\documentclass[12pt,letterpaper]{article}
\usepackage{graphicx}
\usepackage{accents}
\usepackage{xcolor}

\def\title#1{{\bf\Large #1}}
\def\sec#1{\vspace{1.00\baselineskip} \noindent {\bf\large #1}
\vspace{0.25\baselineskip}}

\def\ga{\gamma}

\def\de{\delta}

\def\ep{\epsilon}

\def\om{\omega}

\def\pa{\parallel}

\def\ph{\phi}
\def\ps{\psi}

\def\si{\sigma}

\def\ta{\tau}
\def\th{\theta}

\def\vr{\vec{r}}
\def\ze{\zeta}

\def\d{\dagger}
\def\<{\langle}
\def\>{\rangle}

\def\tsum{{\textstyle\sum}}

\def\d{\dagger}

\def\diag{{\rm diag}}

\def\vn{\vec{n}}
\def\vu{\vec{u}}
\def\vv{\vec{v}}
\def\st{\accentset{\leftrightarrow}{\si}}
\def\ot{\accentset{\leftrightarrow}{\omega}}

\def\ba{\begin{eqnarray}}
\def\ea{\end{eqnarray}}
\def\be{\begin{equation}}
\def\ee{\end{equation}}

\begin{document}

%\centerline{Notes \hfill 2026/07/23}

\begin{center}

\vspace*{1.0\baselineskip}

\title{How to calculate the Wigner angle}

\vspace{1.0\baselineskip}

C. J. McKinstrie \\

{\it\small Independent Photonics Consultant, Manalapan, NJ 07726}

\vspace{0.50\baselineskip}

M. V. Kozlov \\

{\it\small Center for Preparatory Studies, Nazarbayev University, Astana 010000, Kazakhstan}

\vspace{0.50\baselineskip}

Abstract \\

\vspace{0.50\baselineskip}

\parbox[]{6.5in}{\small  Lorentz transformations in time and two space dimensions consist of boosts and rotations, and combinations thereof. In general, the combination of two boosts is not another boost: It is a boost followed by a rotation. The rotation angle is called the Wigner angle. Although it is straightforward to determine the energy and direction of the combined boost, it is difficult to determine the Wigner angle.
In this article, the vector, matrix and spinor derivations of formulas for the Wigner angle are reviewed, and the underlying mathematics and physics are discussed briefly. Although the derivations are different, the results they produce are equivalent, as they should be. Like many physics problems, if one looks at the problem in the right way, it is not difficult to solve.}

\end{center}

\newpage

\sec{1. Introduction}

The theory of special relativity is an important part of the undergraduate and graduate physics curricula \cite{ein23,gol02,jac99,lan13}, and is required to analyze many phenomena of current interest.
In time and one space dimension, every Lorentz transformation is a boost, transformations always act in the same (or opposite) direction and their composition rules are simple. However, in two (and three) dimensions, transformations are combinations of boosts and rotations, boosts can act in different directions and their composition rules are complicated. 

It is well known that the combination of two parallel (or antiparallel) boosts is another boost (as it is in one dimension), but the combination of two nonparallel boosts is not: It is a boost followed by a rotation.
Although it is straightforward to calculate the energy and direction of the combined  boost, it is difficult to calculate the rotation angle, which is called the Wigner angle. In this article, we discuss three ways to determine the Wigner angle.

In Sec. 2, the standard vector (tensor) formalism is used to represent a boost and analyze the composition of two boosts.
A boost is specified by its (dimensionless) velocity $\vv$, in which case its energy $\ga = 1/(1 - v^2)^{1/2}$, or its momentum $\vu = \ga\vv$, in which case $\ga = (1 + u^2)^{1/2}$. (Time and distance are measured in the same units, so time is really $ct$ and velocity is really $\vv/c$.) Two equivalent formulas for the product tensor are derived. The first formula involves the momenta $\vu_1$ and $\vu_2$ that define the boosts, whereas the second involves the momenta $\vu_{12}$ and $\vu_{21}$ that appear in the product tensor. We explain why the angle between the latter momenta is the Wigner angle and derive a formula for it.

The special orthogonal group SO(1,2) consists of real $3 \times 3$ matrices $L$ that satisfy the indefinite orthogonality (Lorentz) condition $L^tSL = S$, where $S = \diag(1,-1,-1)$ is the structure (metric) matrix. This condition ensures that Lorentz transformations of the coordinate vector $T = [t, x, y]^t$ preserve the squared interval $T^tST = t^2 - x^2 - y^2$.
Lorentz matrices represent boosts in the $x$ and $y$ directions, rotations about the $t$ axis (in the $xy$ plane) and combinations of these transformations.
The Lorentz condition imposes significant constraints on Lorentz matrices, which ensure that each matrix has the decomposition $L(\ga, \th_1, \th_2) = R(\th_2)B(\ga)R^t(\th_1)$, where $B(\ga)$ represents a boost in the $x$ direction with energy $\ga$, and $R(\th_i)$ represents a rotation through the (input or output) angle $\th_i$. This decomposition can be rewritten in the form $L(\ga, \th_1, \th_2) = R(\th_{21})B(\ga,\th_1)$, where $B(\ga,\th_1) = R(\th_1)B(\ga)R^t(\th_1)$ represents a boost with direction angle $\th_1$ and $R(\th_{21})$ represents a rotation through the difference (Wigner) angle $\th_{21} = \th_2 - \th_1$. (The second decomposition was mentioned above.)
In Sec. 3, we use the decomposition formula to efficiently derive equivalent formulas for the Wigner angle.

The special unitary group SU(1,1) consists of complex $2 \times 2$ matrices $M$ that satisfy the indefinite unitarity condition $M^\d SM = S$, where $S = \diag(1,-1)$ is the metric matrix. It is well known that SU(1,1) is locally isomorphic to (has the same local structure as) SO(1,2). Each fundamental indefinite-unitary transformation (a real boost, a complex boost and a differential phase shift) corresponds to a Lorentz transformation (a boost in the $x$ direction, a boost in the $y$ direction and a rotation in the $xy$ plane). One can use these relations to derive matrix-product rules for SU(1,1) and deduce the corresponding rules for SO(1,2). In principle, this indirect method (which is often called the spinor method) is simpler than the direct methods, because it involves $2\times 2$ matrices, rather than $3 \times 3$ matrices (tensors). In Sec. 4, we use it to derive formulas for the Wigner half- and full-angles.

Finally, in Sec. 5, the main results of this article are summarized, and the advantages and disadvantages of each approach are discussed briefly. For completeness, the composition of two arbitrary transformations is discussed in the appendix.

\newpage

\sec{2. Vector formalism}

A Lorentz boost is specified by its (dimensionless) momentum $\vu$ and energy $\ga = (1 + u^2)^{1/2}$. Let $t$ and $\vr$ represent time and position, respectively. Then the boost equations are
\ba t' &= &\ga t + \vu\cdot\vr, \label{2.1} \\
\vr' &= &\vu t + [1 + (\ga - 1)\vu\vu\cdot/u^2]\vr \nonumber \\
&= &\vu t + (1 + \ep\vu\vu\cdot)\vr, \label{2.2} \ea
where the unprimed (primed) variables are inputs (outputs) and $\ep = 1/(\ga + 1)$ \cite{gol02,jac99}. These equations represent a transformation from the laboratory frame to a frame that is moving with momentum $-\vu$.
In Eq. (\ref{2.1}), time is mixed with the component of $\vr$ that is parallel to $\vu$. In Eq. (\ref{2.2}), the parallel part $\vr_\pa = \vu\vu\cdot\vr/u^2$ is mixed with time, whereas the perperdicular part $\vr_\perp = (1 - \vu\vu\cdot/u^2)\vr$ is preserved.

One can rewrite Eqs. (\ref{2.1}) and (\ref{2.2}) in the tensor (matrix-like) form
\be T' = BT, \label{2.3} \ee
where $T = (t,\vr)$ and $T' = (t',\vr')$ are coordinate three-vectors and the boost operator
\be B(\vu) = \left[\begin{array}{c|c} \ga & \vu\cdot \\ \hline \vu & 1 + \ep\vu\vu\cdot \end{array}\right]. \label{2.4} \ee
The lines on the right side of Eq. (\ref{2.4}) divide the operator into blocks. The top-left entry acts on a scalar to produce another scalar, the top-right entry acts on a vector to produce a scalar, the bottom-left entry acts on a scalar to produce a vector and the bottom-right entry acts on a vector to produce another vector. Notice that the boost operator is symmetric. 

Now let $E_p = (\ga_p, \vu_p)$ be the energy--momentum three-vector of a particle of unit mass, where $\ga_p^2 - u_p^2 = 1$. This vector transforms in the same way as the coordinate vector ($E_p' = BE_p$). If the particle is at rest before the boost, then $\ga_p = 1$ and $\vu_p = 0$. After the boost,
\be \ga_p' = \ga, \ \ \vu_p' = \vu. \label{2.5} \ee
Thus, the energy and momentum of a particle boosted from rest equal the energy and momentum that defined the boost, respectively (whence the name boost).

Now consider the boosting of a particle that is already moving.
It follows from Eqs. (\ref{2.3}) and (\ref{2.4}), with $(\ga, \vu) = (\ga_2, \vu_2)$ and $(t, \vr) = (\ga_1, \vu_1)$, that
\ba \ga_{21} &= &\ga_2\ga_1 + \vu_2\cdot\vu_1, \label{2.6} \\
\vu_{21} &= &\vu_2\ga_1 + [1 + \ep_2\vu_2\vu_2\cdot] \vu_1 \nonumber \\
&= &\vu_1 + (\ga_1 + \ep_2\vu_2\cdot\vu_1)\vu_2. \label{2.7} \ea
It is tedious, but straightforward, to show that $\ga_{21}^2 - u_{21}^2 = 1$. Thus, $(\ga_{21}, \vu_{21})$ is also an energy--momentum three-vector.

If we had used $(\ga_1, \vu_1)$ to transform $(\ga_2, \vu_2)$, we would have found that
\ba \ga_{12} &= &\ga_1\ga_2 + \vu_1\cdot\vu_2, \label{2.8} \\
\vu_{12} &= &\vu_2 + (\ga_2 + \ep_1\vu_1\cdot\vu_2)\vu_1, \label{2.9} \ea
where $\ga_{12}^2 - u_{12}^2 = 1$. Although $\vu_{12}$ and $\vu_{21}$ are different vectors (and the formulas for them in terms of $\vu_1$ and $\vu_2$ are complicated), their lengths are equal because their associated energies are equal ($\ga_{12} = \ga_{21} = \ga$). Only their directions are different, which means that one vector is a rotated version of the other (in the plane defined by $\vu_1$ and $\vu_2$).

Now consider two successive transformations, specifed by the three-vectors $(\ga_1, \vu_1)$ and $(\ga_2, \vu_2)$. Then, by applying Eqs. (\ref{2.3}) and (\ref{2.4}) twice, one finds that $T'' = B_2B_1T$, where the product operator
\be B_2(\vu_2)B_1(\vu_1) = \left[\begin{array}{c|c} \ga & \vu_{12}\cdot \\ \hline \vu_{21} & \st_{21} \end{array}\right]. \label{2.11} \ee
The consituents of this operator are
\ba \ga &= &\ga_2\ga_1 + \vu_2\cdot\vu_1, \label{2.12} \\
\vu_{12} &= &\vu_2 + (\ga_2 + \ep_1\vu_2\cdot\vu_1)\vu_1, \label{2.13} \\
\vu_{21} &= &\vu_1 + (\ga_1 + \ep_2\vu_2\cdot\vu_1)\vu_2, \label{2.14} \\
\st_{21} &= &1 + \ep_2\vu_2\vu_2\cdot +\ \ep_1\vu_1\vu_1\cdot +\ (1 + \ep_2\ep_1\vu_2\cdot\vu_1)\vu_2\vu_1\cdot. \label{2.15} \ea
If the operator $B_2B_1$ were to act on the three-vector $(1, 0)$, it would produce the three-vector $(\ga_{21}, \vu_{21})$, and if the operator $B_1B_2$ were to act on the three-vector $(1, 0)$, it would produce the three-vector $(\ga_{12}, \vu_{12})$, where $\ga_{12} = \ga_{21} = \ga$. This observation establishes the physical significances of the scalar in Eq. (\ref{2.12}), and the vectors in Eqs. (\ref{2.13}) and (\ref{2.14}). However, the physical significance of the dyadic in Eq. (\ref{2.15}) is not obvious. Notice that this dyadic depends quadratically on $\vu_2$ and $\vu_1$, and is almost symmetric: The only term that depends on the order of the subscripts (boosts) is the last one, which is proportional to $\vu_2\vu_1$. Because of this asymmetry, the composition of two boosts is not another boost (unless $\vu_2$ is parallel or antiparallel to $\vu_1$).

Evidently, Lorentz transformations are not limited to boosts. To ascertain what else they are, consider two pairs of coordinate axes, one in the laboratory frame (LF) and the other in a first moving frame (MF1), which moves with momentum $\vu_1$ relative to the LF. At some reference time, the two origins coincide. How does one transform the LF axes so that they always coincide with the MF1 axes? First, one boosts the LF axes so that their origin coincides (keeps up) with the origin of the MF1 axes. Second (if necessary), one rotates the LF axes so that they align with the MF1 axes. Thus, the most general transformation between the frames consists of a boost followed by a rotation \cite{wig39,cus67}. 
Now consider another set of axes in a second moving frame (MF2), which moves with momentum $\vu_2$ relative to MF1. Then one can transform the MF1 axes so that they coincide with the MF2 axes by boosting and rotating them. But the MF2 axes also move relative to the LF axes, so one can also transform the LF axes to the MF2 axes by boosting and rotating them (in ways that remain to be determined). Thus, Lorentz transformations, and combinations of them, consist of boosts followed by rotations.

The product operator (\ref{2.11}) is asymmetric, so it must represent a boost followed by a rotation, but which ones? Let $\ot_{21}$ be the rotation operator that converts $\vu_{12}$ to $\vu_{21}$ and consider the ansatz
\ba L_{21} &= &\left[\begin{array}{c|c} 1 & 0 \\ \hline 0 & \ot_{21} \end{array}\right]
\left[\begin{array}{c|c} \ga & \vu_{12}\cdot \\ \hline \vu_{12} & 1 + \ep\vu_{12}\vu_{12}\cdot \end{array}\right] \nonumber \\
&= &\left[\begin{array}{c|c} \ga & \vu_{12}\cdot \\ \hline \vu_{21} & \ot_{21} +\ \ep\vu_{21}\vu_{12}\cdot \end{array}\right]. \label{2.21} \ea
Notice that the rotation operator only modifies the bottom entries of the boost operator on which it acts. The requirement that the top entries in operator ({\ref{2.21}) equal the top entries in operator (\ref{2.11}) forces the boost operator to be the one specified by $\vu_{12}$. The bottom left entry equals $\vu_{21}$ by construction. By comparing the bottom right entries, one finds that
\be \st_{21} = \ot_{21} + \ep\vu_{21}\vu_{12}\cdot. \label{2.22} \ee
This entry contains the rotation operator itself, and the dyadic $\vu_{21}\vu_{12}$, both of which are asymmetric.

It only remains to write the rotation operator in terms of vectors.
Let $\vn_1$ and $\vn_2$ be arbitrary unit vectors, let $\th_{21}$ be the angle between them and consider the ansatz
\be \ot_{21} = (\vn_1\cdot\vn_2) + (\vn_1\times\vn_2)\times, \label{2.23} \ee
where the dot product has magnitude $\cos(\th_{21})$ and the cross product has magnitude $\sin(\th_{21})$. It is easy to verify that $\ot_{21}\vn_1 = \vn_2$. The dot product shortens $\vn_1$ and the cross product adds the perpendicular part required to convert it to $\vn_2$.
In the context of transformation (\ref{2.21}),
\be \ot_{21} = {\vu_{12}\cdot\vu_{21} + (\vu_{12}\times\vu_{21})\times \over \ga^2 - 1}, \label{2.24} \ee
because $|\vu_{12}| = |\vu_{21}| = (\ga^2 - 1)^{1/2}$. By combining Eqs. (\ref{2.22}) and (\ref{2.24}), one obtains the operator
\be \st_{21} = {\vu_{12}\cdot\vu_{21} + (\vu_{12}\times\vu_{21})\times \over \ga^2 - 1}
+ {\vu_{21}\vu_{12}\cdot \over \ga + 1}. \label{2.25} \ee
Formula (\ref{2.15}) involves the momenta $\vu_1$ and $\vu_2$ that define the constituent boosts, whereas formula (\ref{2.25}) involves the momenta $\vu_{12}$ and $\vu_{21}$ that appear in the product operator. We demonstrated their equivalence in \cite{mck25b}.

It follows from the first line of Eq. (\ref{2.21}) that the rotation angle, which is called the Wigner angle, is the angle between $\vu_{12}$ and $\vu_{21}$. By combining Eqs. (\ref{2.7}) and (\ref{2.9}), one finds~that
\be \vu_{12}\times\vu_{21} = (b_{12}b_{21} - 1)\vu_1\times\vu_2, \label{2.10} \ee
where $b_{12} = \ga_2 + \ep_1\vu_1\cdot\vu_2$ and $b_{21} = \ga_1 + \ep_2\vu_1\cdot\vu_2$. Notice that the right side of Eq. (\ref{2.10}) is proportional to the sine of the angle between $\vu_1$ and $\vu_2$ ($\th_{21}$), whereas the left side is proportional to the sine of the angle between $\vu_{12}$ and $\vu_{21}$ ($\th_w$). If the former vectors are not parallel, then neither are the latter vectors.
It follows from Eq. (\ref{2.10}) that
\be \sin(\th_w) = {(b_{12}b_{21} - 1)u_2u_1s_{21} \over \ga^2 - 1}, \label{2.31} \ee
where $s_{21} = \sin(\th_{21})$. Although Eq. (\ref{2.31}) looks complicated, the numerator and denominator are both divisible by $\ga - 1$ [Eq. (\ref{2.12})]. The numerator is proportional to
\ba &&(b_{12}b_{21} - 1)u_2u_1 \nonumber \\
&= &(\ga_2u_1 + u_2\de_1c_{21})(u_2\ga_1 + \de_2u_1c_{21}) - u_2u_1 \nonumber \\
&= &(\ga_2\ga_1 - 1)u_2u_1 + (\ga_2\de_2u_1^2 + u_2^2\ga_1\de_1)c_{21} + \de_2\de_1u_2u_1c_{21}^2,
\label{2.32} \ea
where $c_{21} = \cos(\th_{21})$ and $\de_i = \ga_i - 1$.
The first term in Eq. (\ref{2.32}) is the product of $\ga_2\ga_1 - 1$ and $u_2u_1$, and the last term is the product of $u_2u_1c_{21}$ and $\de_2\de_1c_{21}$, so consider the ansatz
\ba &&(\ga_2\ga_1 - 1 + u_2u_1c_{21})(u_2u_1 + \de_2\de_1c_{21}) \nonumber \\
&= &(\ga_2\ga_1 - 1)u_2u_1 + [(\ga_2\ga_1 - 1)\de_2\de_1 + u_2^2u_1^2]c_{21} + \de_2\de_1u_2u_1c_{21}^2. \label{2.33} \ea
In Eqs. (\ref{2.32}) and (\ref{2.33}), the coefficients of $c_{21}$ both equal $(2\ga_2\ga_1 + \ga_2 + \ga_1)\de_2\de_1$, so the latter equation is the required factorization of the former. By combining the preceding results, one finds that the Wigner angle is specified implicitly by the equation
\be \sin(\th_w) = {(u_2u_1 + \de_2\de_1c_{21})s_{21} \over \ga_2\ga_1 + 1 + u_2u_1c_{21}}. \label{2.34} \ee
%.
Notice that $\th_w$ only depends on the difference angle $\th_{21}$.

The derivation of Eq. (\ref{2.34}) was based on the cross product of $\vu_{12}$ and $\vu_{21}$. In \cite{mck25b}, we used the dot product to derive the alternative equation
\be \cos(\th_w) = {\ga_2 + \ga_1 + (u_2u_1 + \de_2\de_1c_{21})c_{21} \over \ga_2\ga_1 + 1 + u_2u_1c_{21}}. \label{2.35} \ee
We also checked that formulas (\ref{2.34}) and (\ref{2.35}) are consistent ($c_w^2 + s_w^2 = 1$). Notice that their common denominator is $\ga + 1$ [Eq. (\ref{2.12})].

It is convenient to define the transformation parameter
\be \ta = u_2u_1/\de_2\de_1 = [(\ga_2 + 1)(\ga_1 + 1)/(\ga_2 - 1)(\ga_1 - 1)]^{1/2}, \label{2.41} \ee
which should not be confused with the proper time. By multiplying the numerators and denominators of Eqs. (\ref{2.34}) and (\ref{2.35}) by $2/\de_2\de_1$, and using the identities
\ba \ta^2 + 1 &= &2(\ga_2\ga_1 + 1)/\de_2\de_1, \label{2.42} \\
\ta^2 - 1 &= &2(\ga_2 + \ga_1)/\de_2\de_1, \label{2.43} \ea
one finds that
\ba s_w &= &{2(u_2u_1/\de_2\de_1 + c)s \over 2(\ga_2\ga_1 + 1)/\de_2\de_1 + 2(u_2u_1/\de_2\de_1)c}
\nonumber \\
&= &{2(\ta + c)s \over \ta^2 + 1 + 2\ta c} \ = \ {2(\ta + c)s \over (\ta + c)^2 + s^2}, \label{2.44} \\
c_w &= &{2(\ga_2 + \ga_1)/\de_2\de_1 + 2(u_2u_1/\de_2\de_1 + c)c \over 2(\ga_2\ga_1 + 1)/\de_2\de_1 + 2(u_2u_1/\de_2\de_1)c} \nonumber \\
&= &{\ta^2 - 1 + 2(\ta + c)c \over \ta^2 + 1 + 2\ta c} \ = \ {(\ta + c)^2 - s^2 \over (\ta + c)^2 + s^2}, \label{2.45} \ea
where $c$ and $s$ are abbreviations of $c_{21}$ and $s_{21}$, respectively.
Equations (\ref{2.44}) and (\ref{2.45}) are equivalent to Eqs. (13) and (10) of \cite{ben85}, respectively.

An advantage of the vector formalism is that it generalizes automatically from two to three space dimensions. The (input) vectors $\vu_1$ and $\vu_2$, which appear in the boost tensors, specify a plane in three-space. The (output) vectors $\vu_{12}$ and $\vu_{21}$, which appear in the product tensor, are linear combinations of $\vu_1$ and $\vu_2$, so they lie in the aforementioned plane. The vector cross- and dot-products are independent of the coordinate systems used to represent them. If one aligns the $x$ axis with $\vu_1$, the $y$ axis with $\vu_2 - (\vu_2\cdot\vu_1)\vu_1/u_1^2$ and the $z$ axis with $\vu_1\times\vu_2$, then the transformation is effectively two-dimensional, and Eqs. (\ref{2.34}) and (\ref{2.35}) still apply. Two caveats are worth mentioning: First, Eq. (\ref{2.23}) can be rewritten in the form $\ot_{21} = c_{21} + s_{21}\vn\times$, where $\vn = (\vn_1\times\vn_2)/|\vn_1\times\vn_2|$ is the unit vector that defines the axis of rotation. In three dimensions, it remains true that $\ot_{21}\vn_1 = \vn_2$. Nonetheless, one should add to the rotation tensor the term $(1 - c_{21})\vn\vn\cdot$, which allows the tensor to preserve the parallel components of the vectors on which it acts \cite{mck25c}. Second, in two dimensions, the third (successive) boost is coplanar (in the plane defined by the first two boosts), whereas in three dimensions, the third boost need not be coplanar.

\newpage

\sec{3. Matrix formalism}

Let $T = [t, x, y]^t$ and $T' = [t', x', y']^t$ be $3 \times 1$ coordinate vectors, and let $L$ be a real $3 \times 3$ matrix. Then a Lorentz transformation can be written in the matrix form
\be T' = LT, \label{3.1} \ee
provided that the transformation matrix satisfies the equivalent equations
\be L^tSL = S, \ \ L^{-1} = SL^tS, \label{3.2} \ee
where $S = \diag(1, -1,-1)$ is the structure (metric) matrix. The first of Eqs. (\ref{3.2}) is called the Lorentz condition. It ensures that the inner product (squared interval) $T^tST = t^2 - x^2 - y^2$ is conserved. The set of Lorentz matrices with determinant 1 (rather than $-1$) forms a group under multiplication \cite{mck25b,mck25c}. It is called the special orthogonal group SO(1,2), or the indefinite orthogonal group, because the metric matrix is indefinite (has positive and negative eigenvalues, which allow the squared interval to be positive or negative). The set of coordinate vectors forms a vector space under addition, which is called Minkowski space.

Examples of Lorentz matrices include the boost matrices
\be B_x(\ga) = \left[\begin{array}{ccc} \ga & u & 0 \\ u & \ga & 0 \\ 0 & 0 & 1 \end{array}\right], \ \ 
B_y(\ga) = \left[\begin{array}{ccc} \ga & 0 & u \\ 0 & 1 & 0 \\ u & 0 & \ga \end{array}\right], \label{3.3} \ee
where $\ga$ and $u$ are the (dimensionless) energy and momentum parameters, respectively. The Lorentz and determinant conditions both require that $\ga^2 - u^2 = 1$, which allows one to write $\ga = \cosh(\ze)$ and $u = \sinh(\ze)$, where $\ze$ is the boost parameter (rapidity). The rotation matrix
\be R(\th) = \left[\begin{array}{ccc} 1 & 0 &0 \\ 0 & c & -s \\ 0 & s & c \end{array}\right], \label{3.4} \ee
where $\th$ is the rotation angle, $c = \cos(\th)$ and $s = \sin(\th)$.
SO(1,2) is closed under multiplication, so products of Lorentz matrices are also Lorentz matrices. Consider the product
\be B(\ga, \th) = R(\th)B_x(\ga)R^t(\th), \label{3.5} \ee
which is symmetric. Written explicitly,
\be B(\ga,\th) = \left[\begin{array}{ccc} \ga & uc & us \\ uc & 1 + \de c^2 & \de cs \\ us & \de sc & 1 + \de s^2 \end{array}\right], \label{3.6} \ee
where $\de = \ga - 1$. In Eq. (\ref{3.5}), $R^t$ aligns the $x$ axis with the intended boost direction, $B_x$ effects the boost and $R$ returns the axes to their original orientations, so matrix (\ref{3.6}) represents a boost of energy $\ga$ with direction angle $\th$. For example, $B_y(\ga) = R(\pi/2)B_x(\ga)R^t(\pi/2)$. Equation (\ref{3.5}) is a similarity transformation of $B_x$, which preserves the boost energy. If one were to rewrite matrix (\ref{3.6}) in terms of momentum components ($c = u_x/u$ and $s = u_y/u$), one would obtain tensor (\ref{2.4}).

The Lorentz condition imposes significant constraints on Lorentz matrices.
In \cite{mck25b,mck25c} we explained why every Lorentz matrix can be written in the block forms
\ba L &= &\left[\begin{array}{cc} \ga & R \\ C & N + \ep CR \end{array}\right] \nonumber \\
&= &\left[\begin{array}{cc} 1 & 0 \\ 0 & N \end{array}\right]
\left[\begin{array}{cc} \ga & R \\ R^t & I + \ep R^tR \end{array}\right] \nonumber \\
&= &\left[\begin{array}{cc} \ga & C^t \\ C & I + \ep CC^t \end{array}\right]
\left[\begin{array}{cc} 1 & 0 \\ 0 & N \end{array}\right], \label{3.11} \ea
where $\ga$ and $\ep = 1/(\ga + 1)$ are scalars, $R$ (temporarily) and $C$ are row and column vectors, respectively, and $N$ is a rotation matrix ($C = NR^t$, so $C^tN = R$). The matrices that involve $\ga$ and $R$, or $\ga$ and $C$, are symmetric, so they represent boosts in directions that are determined by $R$ or $C$. Thus, every Lorentz transformation can be written as a boost followed (or preceded) by a rotation. Notice that Eqs. (\ref{2.21}) and Eq. (\ref{3.11}) are equivalent. The derivation of the former equation was based on physics, whereas the derivation of the latter is based on mathematics.

In terms of components,
\ba L &= &\left[\begin{array}{ccc} \ga & uc_1 & us_1 \\ uc_2 & c_{21} + \de c_2c_1 & -s_{21} + \de c_2s_1 \\
us_2 & s_{21} + \de s_2c_1 & c_{21} + \de s_2s_1 \end{array}\right] \nonumber \\
&= &\left[\begin{array}{ccc} 1 & 0 & 0 \\ 0 & c_2 & -s_2 \\
0 & s_2 & c_2 \end{array}\right]
\left[\begin{array}{ccc} \ga & u & 0 \\ u & \ga & 0 \\ 0 & 0 & 1 \end{array}\right]
\left[\begin{array}{ccc} 1 & 0 & 0 \\ 0 & c_1 & s_1 \\
0 & -s_1 & c_1 \end{array}\right], \label{3.12} \ea
where $c_i = \cos(\th_i)$, $s = \sin(\th_i)$, and $\th_{21} = \th_2 - \th_1$. 
By comparing the first lines of Eqs. (\ref{3.11}) and (\ref{3.12}), one finds that $R = u[c_1, s_1]$ and $C = u[c_2, s_2]^t$ are momentum vectors with (common) length $u$, and direction angles $\th_1$ and $\th_2$.
It follows from the second line of Eq. (\ref{3.12}) that every Lorentz matrix has the decomposition
\be L(\ga, \th_1, \th_1) = R(\th_2)B(\ga)R(\th_1)^t, \label{3.13} \ee
in which $B = B_x$ is the fundamental boost matrix and $R_i = R(\th_i)$ is a two-dimensional rotation matrix \cite{mck25b,mck25c}. Consequently, $\th_1$ and $\th_2$ are called the input and output angles, respectively, and $\th_{21}$ is called the difference angle. Notice that Eq. (\ref{3.13}) is a simple generalization of Eq. (\ref{3.5}). It shows that a general Lorentz matrix is specified by the independent (free) parameters $\ga$, $\th_1$ and $\th_2$.

Every real matrix $M$ has the singular-value (Schmidt) decomposition $M = QDP^t$, where $D$ is a nonnegative diagonal matrix, which represents dilations, and $P$ and $Q$ are orthogonal matrices, which represent basis changes \cite{hor13}. In Eq. (\ref{3.13}), the dilation matrix is replaced by a boost matrix, and three-dimensional orthogonal transformations are replaced by two-dimensional rotations (which are simpler). In the context of Lorentz transformations, the Schmidt-like decomposition (\ref{3.13}) is even more useful than the Schmidt decomposition.

Let $L = [l_{ij}]$, where $i$ and $j$ vary from 0 to 2. Then it follows from the first line of Eq. (\ref{3.12}) that the energy
\be \ga = \l_{00}. \label{3.14} \ee
The input and output angles are specified implicitly by the equations
\be \tan(\th_1) = l_{02}/l_{01}, \ \ 
\tan(\th_2) = l_{20}/l_{10}, \label{3.15} \ee
and the difference angle is specified by the equations
\ba (\ga + 1)\cos(\th_{21}) &= &l_{11} + l_{22}, \label{3.16} \\
(\ga + 1)\sin(\th_{21}) &= &l_{21} - l_{12}, \label{3.17} \ea
from which it follows that
\be \tan(\th_{21}) = (l_{21} - l_{12})/(l_{11} + l_{22}). \label{3.18} \ee
If one is presented with a Lorentz matrix (a boost or rotation matrix, or a combination of such matrices), one can read off the energy, and determine the input, output and difference angles, with little effort.

Before proceeding further, two remarks are in order. First, (in two dimensions) rotation matrices satisfy the product rule $R(\th_2)R(\th_1) = R(\th_2 + \th_1)$. It follows from this rule and Eq. (\ref{3.13}) that
\be R(\th_3) L(\ga, \th_1, \th_2) R^t(\th_3) = L(\ga, \th_1 + \th_3, \th_2 + \th_3). \label{3.19} \ee
The similarity transform of a Lorentz matrix is another Lorentz matrix. This transformation changes the input and output angles in simple ways, but does not change the difference angle or the energy. Notice that if $\th_3 = -\th_1$, the transformed matrix depends only on the difference angle $\th_{21}$.

Second, consider the transformation equation $Y = LX$, where (temporarily) $X$ and $Y$ play the roles of $T$ and $T'$, respectively. If one decides to work in terms of the rotated vectors $X' = R^tX$ and $Y' = R^tY$, then $Y' = L'X'$, where $L' = R^tLR$. The inverse of the last equation is $L = RL'R^t$.

Now consider the composition of two arbitrary boosts, which are specified by the energies $\ga_1$ and $\ga_2$, and the direction angles $\th_1$ and $\th_2$.  In a (rotated) frame that is aligned with the first boost, the first boost angle is 0 and the second boost angle is the difference angle $\th_{21}$. The product matrix
\ba B'_2B'_1 &= &\left[\begin{array}{ccc} \ga_2 & u_2c & u_2s \\ u_2c & 1 + \de_2c^2 & \de_2cs \\
u_2s & \de_2cs & 1 + \de_2s^2 \end{array}\right]
\left[\begin{array}{ccc} \ga_1 & u_1 & 0 \\u_1 & \ga_1 &0 \\
0 & 0 & 1 \end{array}\right] \nonumber \\
&= &\left[\begin{array}{ccc} \ga_2\ga_1 + u_2u_1c & \ga_2u_1 + u_2\ga_1c & u_2 s \\
(u_2\ga_1 + \de_2u_1c)c + u_1 & (u_2u_1 + \de_2\ga_1c)c + \ga_1 & \de_2cs \\
(u_2\ga_1 + \de_2u_1c)s & (u_2u_1 + \de_2\ga_1c)s & 1 + \de_2s^2 \end{array}\right], \label{3.21} \ea
where $c$ and $s$ are abbreviations of $c_{21}$ and $s_{21}$, respectively. It follows from Eqs. (\ref{3.14}) and (\ref{3.21}) that the composite energy
\be \ga' = \ga_2\ga_1 + u_2u_1c. \label{3.22} \ee
It follows from Eqs. (\ref{3.15}) and (\ref{3.21}) that the composite input and output angles are specified by the equations
\be \tan(\ph') = {u_2s \over \ga_2u_1 + u_2\ga_1c}, \ \ 
\tan(\th') = {(u_2\ga_1 + \de_2u_1c)s \over u_1 + (u_2\ga_1 + \de_2u_1c)c}, \label{3.24} \ee
respectively. (When discussing product matrices, it is easier to use different symbols for the input and output angles.)
In the context of combined boosts, the difference angle ($\th' - \ph'$) is called the Wigner angle ($\th'_w$).
It follows from Eqs. (\ref{3.18}) and (\ref{3.21}) that the Wigner angle is specified by the equation
\be \tan(\th'_w) = {(u_2u_1 + \de_2\de_1c)s \over \ga_2 + \ga_1 + (u_2u_1 + \de_2\de_1c)c}. \label{3.25} \ee

According to Eq. (\ref{3.19}), in the (unrotated) laboratory frame, the input and output angles
\be \ph = \ph' + \ph_1, \ \ \th = \th' + \ph_1, \label{3.26} \ee
respectively. Backward rotation does not change the energy or difference angle, so Eq. (\ref{3.22}) is the well-known formula for the combined energy $\ga$ and Eq. (\ref{3.25}) is the sought-after formula for the Wigner angle $\th_w$. It is consistent with Eqs. (\ref{2.34}) and (\ref{2.35}).

Now consider Lorentz transformations in time and three space dimensions.
The special orthogonal group SO(1,3) is the set of real $4 \times 4$ matrices $L$ that satisfy the Lorentz condition $L^tSL = S$, where $S = \diag(1,-1,-1,-1)$, and have determinant 1. SO(1,3) has six fundamental matrices, which represent boosts along the $x$, $y$ and $z$ axes, and rotations about the $x$, $y$ and $z$ axes.
Every Lorentz matrix can be written in the form of matrix (\ref{3.11}), in which $C$ is a $3 \times 1$ column vector, $R$ is a $1 \times 3$ row vector and $N$ is the $3 \times 3$ rotation matrix that converts (rotates) $R^t$ to $C$. This matrix involves two polar angles, which specify the axis of rotation, and one rotation angle, which is the analog of the Wigner angle.

It is easy to verify that
\ba \bar{N}_2L\bar{N}_1^t &= &\left[\begin{array}{cc} 1 & 0 \\ 0 & N_2 \end{array}\right]
\left[\begin{array}{cc} \ga & R \\ C & N + \ep CR \end{array}\right]
\left[\begin{array}{cc} 1 & 0 \\ 0 & N_1^t \end{array}\right] \nonumber \\
&= &\left[\begin{array}{cc} \ga & RN_1^t \\ N_2C & N_2NN_1^t + \ep N_2CRN_1^t \end{array}\right] \nonumber \\
&= &\left[\begin{array}{cc} \ga & R_1 \\ C_2 & N_{21} + \ep C_2R_1 \end{array}\right], \label{3.31} \ea
where $C_2 = N_2C$ is the new column vector, $R_1 = RN_1^t = (N_1R^t)^t$ is the new row vector and $N_{21} = N_2NN_1^t$ is the new rotation matrix. (Although rotation matrices do not commute in three dimensions, it remains true that the product of rotation matrices is another rotation matrix.) It follows from the identity $C = NR^t$ that $C_2 = N_2N(N_1^tN_1)R^t = (N_2NN_1^t)R_1^t$, so $N_{21}$ is the matrix that rotates $R_1^t$ to $C_2$. Thus, the generalized similarity transform of a general Lorentz matrix is another general Lorentz matrix. This transformation changes the row and column momentum vectors in simple ways, but does not change the energy. The relation $N \rightarrow N_2NN_1^t$ is the three-dimensional analog of the two-dimensional relation $\th_w \rightarrow \th_w + \th_2 - \ph_1$ [Eqs. (\ref{a11}) and (\ref{a27})]. If the input matrix $L = B_x$ represents a boost in the $x$ direction ($C^t = R = [1, 0, 0]$), then the output matrix is a general matrix [Eq. (\ref{3.11})] and Eq. (\ref{3.31}) is its Schmidt-like decomposition [Eq. (\ref{3.13})].

\newpage

\sec{4. SU(1,1) and SO(1,2)}

In this section, we introduce the special unitary group SU(1,1) and demonstrate the local isomorphism between SU(1,1) and SO(1,2), which is known. This structural similarity allows one to derive results for SU(1,1) and deduce the corresponding results for SO(1,2), without further effort.
Before proceeding with this demonstration, it is helpful to describe a similar relation between SU(2) and SO(3), which is better known.

The special unitary group SU(2) is the set of complex $2 \times 2$ matrices $U$ that satisfy the unitarity condition $U^\d U = I$. If $A$ is a complex $2 \times 1$ vector on which $U$ acts, then the unitarity condition ensures that the inner product (norm) $A^\d A$ is conserved. In polarization optics \cite{gor00,gol11,chi18}, $A = [A_x, A_y]^t$ is the amplitude vector of a wave and the norm is its power (photon flux). The set of amplitude vectors forms a vector space under addition, which is called Jones space.

The special orthogonal group SO(3) is the set of real $3 \times 3$ matrices $R$ that satisfy the orthogonality condition $R^tR = I$. If $X$ is a real $3 \times 1$ vector on which $R$ acts, then the orthogonality condition ensures that the norm $X^tX$ is conserved. In three-dimensional rotations, the vector $X = [x, y, z]$ is the position vector and the norm is its squared length. The set of position vectors forms a vector space, which (in relation to polarization optics) is called Stokes space.

Polarization transformations are locally isomorphic to (have the same local stucture as) three-dimensional rotations \cite{mck25c,gor00}. The fundamental transformations in Jones space (real and complex beam-splitter transformations, and a differential phase shift) correspond to fundamental transformations in Stokes space (rotations about the $x$, $y$ and $z$ axes). One can use the correspondence between the associated fundamental matrices to derive results (including product rules) for polarization transformations and deduce the corresponding results for rotations. For SU(2) and SO(3), this local-isomorphism method simplifies the calculations significantly \cite{mck25c,rui13}, which is why it is well known and often used.

The main subject of this section, which is the local isomorphism between SU(1,1) and SO(1,2), was discussed in \cite{pen84,mis17,tor23a,tor23b}.
The authors' analyses are mathematical, but clear, so it would be pointless to repeat them here. Instead, we will take an empirical approach and demonstrate the local isomorphism directly.

The special unitary group SU(1,1) is the set of complex $2 \times 2$ matrices $M$ that satisfy the equivalent conditions
\be M^\d SM = S,  \ \ M^{-1} = SM^\d S, \label{4.1} \ee
where the metric matrix $S = \diag(1,-1)$. They also satisfy the (special) condition $\det(M) = 1$. The first of Eqs. (\ref{4.1}) is called the indefinite unitarity condition, because the metric matrix is indefinite (has positive and negative components). If $A$ is a complex $2 \times 1$ vector on which $M$ acts, then the indefinite unitarity condition ensures that the generalized inner product $A^\d SA$ is conserved. The set of vectors (spinors) forms a vector space, which is called spinor space \cite{pen84,mis17}. (This space has the same vectors as Jones space, but a different inner product.)

In three-wave mixing \cite{boy20}, which occurs in second-order nonlinear media, a strong pump wave drives weak signal and idler waves ($\pi_p \rightarrow \pi_s + \pi_i$, where $\pi_i$ represents a photon with frequency $\om_i$). Similarly, in four-wave mixing \cite{mar08}, which occurs in third-order nonlinear media, one or two strong pump waves drive weak signal and idler waves ($2\pi_p \rightarrow \pi_s + \pi_i$ or $\pi_p + \pi_q \rightarrow \pi_s + \pi_i$). In each process, signal and idler photons are produced in pairs, so the difference between the signal and idler powers (photon fluxes) is constant. In the strong-pump approximation, the pump power(s) is (are) constant, and the coupled equations for the signal and idler amplitudes are linear. The amplitude vector $A = [A_s, A_i^*]^t$ and the inner product is the signal--idler power (flux) difference.

Examples of indefinite unitary matrices include the boost-like matrices
\be B_1 = \left[\begin{array}{cc} \mu_1 & \nu_1 \\ \nu_1 & \mu_1 \end{array}\right], \ \ 
B_2 = \left[\begin{array}{cc} \mu_2 & i\nu_2 \\ -i\nu_2 & \mu_2 \end{array}\right], \label{4.2} \ee
where $\mu_i$ and $\nu_i$ are real parameters that satisfy the auxiliary equation $\mu_i^2 - \nu_i^2 = 1$, and the differential phase-shift matrix
\be P_3 = \left[\begin{array}{cc} e_3 & 0 \\ 0 & e_3^* \end{array}\right], \label{4.3} \ee
where $e_3 = \exp(i\ph_3)$.
These matrices describe a real boost, a complex boost and a differential phase shift, respectively.
Matrices (\ref{4.2}) are not Lorentz boost matrices, but we use the term because the identity $\mu^2 - \nu^2 = 1$ is equivalent to $\ga^2 - u^2 = 1$. This identity allows one to write $\mu = \cosh(\ze)$ and $\nu = \sinh(\ze)$, where $\ze$ is the boost parameter.

SU(1,1) is closed under multiplication, so the products of indefinite unitary matrices are also indefinite unitary matrices.
For example, consider the product matrix
\ba M(\mu,\ph) &= &P(\ph)B_1(\mu)P^\d(\ph) \nonumber \\
&= &\left[\begin{array}{cc} \mu & \nu e_\ph^2 \\ \nu(e_\ph^*)^2 & \mu \end{array}\right]. \label{4.4} \ea
We will refer to all such matrices as boost matrices, because they are similarity transforms of the real boost matrix. Such transformations do not affect the magnitude or phase of $\mu$.
In the same way that a Lorentz boost in the $y$ direction is a rotated version of a boost in the $x$ direction [Eq. (\ref{3.5}) with $\ph = \pi/2$], the complex boost is a phase-shifted version of the real boost [Eq. (\ref{4.4}) with $\ph = \pi/4$].

By applying the indefinite unitarity condition (\ref{4.1}) to a matrix $M$ and demanding that $\det(M) = 1$ [rather than $\exp(i\ps)$], one can show that every indefinite unitary matrix can be written in the form
\be M = \left[\begin{array}{cc} \mu & \nu \\ \nu^* & \mu^* \end{array}\right], \label{4.5} \ee
where $\mu$ and $\nu$ are (temporarily) complex parameters that satisfy the auxiliary equation $|\mu|^2 - |\nu|^2 = 1$. Matrix (\ref{4.5}) is specified by the free parameters $|\mu|$, $\ph_\mu$ and $\ph_\nu$.
Matrix (\ref{4.4}) has the generalization
\ba M(\mu,\ph_1,\ph_2) &= &P(\ph_2)B_1(\mu)P^\d(\ph_1) \nonumber \\
&= &\left[\begin{array}{cc} \mu e_2e_1^* & \nu e_2e_1 \\ \nu e_2^*e_1^* & \mu e_2^*e_1 \end{array}\right], \label{4.6} \ea
where $e_i = \exp(i\ph_i)$.
By comparing matrices (\ref{4.5}) and (\ref{4.6}), one finds that they have the same form, in which the components in the bottom row are conjugates of components in the top row. The equivalence conditions are
$\mu_c = \mu_re_2e_1^*$ and $\nu_c = \nu_re_2e_1$, where the subscripts $c$ and $r$ stand for complex and real, respectively.
If matrix (\ref{4.5}) is specified, then
\be \mu_r = |\mu_c|, \ \ \ph_1 = (\ph_\nu - \ph_\mu)/2, \ \ \ph_2 = (\ph_\mu + \ph_\nu)/2, \label{4.7} \ee
whereas if matrix (\ref{4.6}) is specified, then
\be |\mu_c| = \mu_r, \ \ \ph_\mu = \ph_2 - \ph_1, \ \ \ph_\nu = \ph_2 + \ph_1. \label{4.8} \ee

The preceding results show that every indefinite unitary matrix has the Schmidt-like decomposition
\be M = P(\ph_2)B(\mu)P^\d(\ph_1),  \label{4.9} \ee
where $B = B_1$ is a real boost matrix and $P_i = P(\ph_i)$ is a phase-shift matrix. The parameters $\mu$, $\ph_1$ and $\ph_2$, which are specified by Eqs. (\ref{4.7}), are called the boost strength, and the input and output phase-angles (phases), respectively. Notice that the difference phase $\ph_2 - \ph_1 = \ph_\mu$. % [Eqs. (\ref{4.8})].
This phase is also called the Wigner phase.
It is easy to verify that
\be P(\ph_3)M(\mu,\ph_1,\ph_2)P^\d(\ph_3) = M(\mu,\ph_1+\ph_3,\ph_2+\ph_3). \label{4.10} \ee
%nothe
The similarity transform of an indefinite unitary matrix is another indefinite unitary matrix. This transformation changes the input and output phases in simple ways, but does not change the difference (Wigner) phase or the boost strength. Notice the similarities between Eqs. (\ref{3.13}) and (\ref{4.9}), and Eqs. (\ref{3.19}) and (\ref{4.10}).

Let $X$ and $Y$ be column vectors, and consider the transformation $Y = MX$. If one decides to work with the phase-shifted vectors $X' = P^\d X$ and $Y' = P^\d Y$, then $Y' = M'X'$, where $M' = P^\d MP$. The inverse of the last equation is $M = PM'P^\d$.

Now consider the composition of two boosts, which are specified by the boost strengths $\mu_1$ and $\mu_2$, and the phases $\ph_1$ and $\ph_2$. and suppose that
\be Y = (P_2B_2P_2^\d)(P_1B_1P_1^\d)X. \label{4.11} \ee
Then, as explained above, this equation can be rewritten as $Y' = M'X'$, where the primed vectors $X'= P_1^\d X$ and $Y' = P_1^\d Y$, and the primed matrix
\be .M' = (P_1^\d P_2)B_2(P_2^\d P_1)B_1. \label{4.12} \ee
The first three matrices on the right side of Eq. (\ref{4.12}) represent a complex boost, with phase $\ph_{21} = \ph_2 - \ph_1$ [Eq. (\ref{4.4})]. Hence, the primed matrix
\ba M' &= &\left[\begin{array}{cc} \mu_2 & \nu_2f_{21} \\ \nu_2f_{21}^* & \mu_2 \end{array}\right]
\left[\begin{array}{cc} \mu_1 & \nu_1 \\ \nu_1 & \mu_1 \end{array}\right], \nonumber \\
&= &\left[\begin{array}{cc} \mu_2\mu_1 + \nu_2\nu_1f_{21} & \mu_2\nu_1 + \nu_2\mu_1f_{21} \\ \mu_2\nu_1 + \nu_2\mu_1f_{21}^* & \mu_2\mu_1 + \nu_2\nu_1f_{21}^* \end{array}\right], \label{4.13} \ea
where $f_{21} = \exp(i2\ph_{21})$. %(Notice the factor of 2.)
It follows from Eqs. (\ref{4.5}) and (\ref{4.13}) that the squared boost parameters
\ba |\mu'|^2 &= &\mu_2^2\mu_1^2 + \nu_2^2\nu_1^2 + 2\mu_2\nu_2\mu_1\nu_1d_{21}, \label{4.14} \\
|\nu'|^2 &= &\mu_2^2\nu_1^2 + \nu_2^2\mu_1^2 + 2\mu_2\nu_2\mu_1\nu_1d_{21}, \label{4.15} \ea
where $d_{21} = \cos(2\th_{21})$. ($d$ stands for double and is the letter that follows $c$.)
It is easy to verify that the difference $|\mu'|^2 - |\nu'|^2 = 1$ (as it should do) and the sum
\be |\mu'|^2 + |\nu'|^2 = (\mu_2^2 + \nu_2^2)(\mu_1^2 + \nu_1^2) + (2\mu_2\nu_2)(2\mu_1\nu_1)d_{21}. \label{4.16} \ee
One can use $|\mu'|^2$, $|\nu'|^2$ or $|\mu'|^2 + |\nu'|^2$ to quantify the boost strength. The component phases are specified implicitly by the equations
\be \tan(\ph_\mu') = {\nu_2\nu_1t_{21} \over \mu_2\mu_1 + \nu_2\nu_1d_{21}}, \ \ 
\tan(\ph_\nu') = {\nu_2\mu_1t_{21} \over \mu_2\nu_1 + \nu_2\mu_1d_{21}}, \label{4.18} \ee
where $t_{21} = \sin(2\ph_{21})$. ($t$ is the letter that follows $s$.)
In the context of combined boosts, it is better to use separate symbols for the input phase $\ph'$ and the output phase $\th'$.
It follows from Eqs. (\ref{4.7}) and (\ref{4.18}) that
\ba \tan(2\ph') %&= &\tan(\ph'_\nu - \ph'_\mu) \nonumber \\
&= &{\mu_2\nu_2t_{21} \over (\mu_2^2 + \nu_2^2)\mu_1\nu_1 + \mu_2\nu_2(\mu_1^2 + \nu_1^2)d_{21}}, \label{4.19} \\
\tan(2\th') %&= &\tan(\ph'_\mu + \ph'_\nu) \nonumber \\
&= &{[\mu_2\nu_2(\mu_1^2 + \nu_1^2) + 2\nu_2^2\mu_1\nu_1d_{21}]t_{21}
\over \mu_1\nu_1 + [\mu_2\nu_2(\mu_1^2 + \nu_1^2) + 2\nu_2^2\mu_1\nu_1d_{21}]d_{21}}. \label{4.20} \ea

The unprimed matrix $M = P_1M'P_1^\d$. Written explicitly,
\be M = \left[\begin{array}{cc} \mu_2\mu_1 + \nu_2\nu_1f_{21} & (\mu_2\nu_1 + \nu_2\mu_1f_{21})e_1^2 \\ (\mu_2\nu_1 + \nu_2\mu_1f_{21}^*)(e_1^*)^2 & \mu_2\mu_1 + \nu_2\nu_1f_{21}^* \end{array}\right]. \label{4.21} \ee
It follows from Eqs. (\ref{4.13}) and (\ref{4.21}) that the component phases are
\be \ph_\mu = \ph'_\mu, \ \ \ph_\nu = \ph'_\nu + 2\ph_1, \label{4.22} \ee
and it follows from Eqs. (\ref{4.7}) and (\ref{4.21}) that the input and output phases are
\be \ph = (\ph'_\nu - \ph'_\mu)/2 +\ph_1, \ \ \th = (\ph'_\mu + \ph'_\nu)/2 + \ph_1, \label{4.23} \ee
respectively.
Notice that the Wigner phase $\th - \ph = \ph'_\mu = \ph_\mu$ does not depend on $\ph_1$ directly. [See the comment after Eq. (\ref{4.10}).] Notice also that this phase is specified by the first of Eqs. (\ref{4.18}), which was obtained by a simple matrix multiplication. This formula is equivalent to formulas derived previously \cite{tor23a,tor23b,van84,rho04}.

The preceding results can be simplified slightly. As explained after Eq. (\ref{4.3}), one can write $\mu = \cosh(\ze) = C$ and $\nu = \sinh(\ze) = S$. It follows from these definitions that $C^2 + S^2 = \cosh(2\ze) = D$ and $2SC = \sinh(2\ze) = T$. By using this notation, one can rewrite Eq. (\ref{4.16}) as
\be D' = D_2D_1 + T_2T_1d_{21}, \label{4.31} \ee
and one can rewrite Eqs. (\ref{4.19}) and (\ref{4.20}) as
\ba \tan(2\phi') &= &{T_2t_{21} \over D_2T_1 + T_2D_1d_{21}}, \label{4.32} \\ 
\tan(2\th') &= &{[T_2D_1 + (D_2 - 1)T_1d_{21}]t_{21} \over T_1 + [T_2D_1 + (D_2 - 1)T_1d_{21}]d_{21}}, \label{4.33} \ea
respectively. Notice that all the arguments in these equations are double arguments.
By applying a trigonmetric identity to the first of Eqs. (\ref{4.18}), and using the identities $C^2 = (D + 1)/2$ and $S^2 = (D - 1)/2$, one can show that
\be \tan(2\ph'_\mu) = {[T_2T_1 + (D_2 - 1)(D_1 - 1)d_{21}]t_{21} \over D_2 + D_1 + [T_2T_1 + (D_2 - 1)(D_1 - 1)d_{21}]d_{21}}. \label{4.34} \ee
If one were to derive Eq. (\ref{4.34}) directly ($\ph'_\mu = \th' - \ph'$), by using Eqs. (\ref{4.32}) and (\ref{4.33}), one would obtain a fraction whose numerator and denominator are both proportional to $\ga' - 1$. By canceling this common factor, one would obtain Eq. (\ref{4.34}).

In passing, the composition of two symplectic dilations was discussed in \cite{mck25a}. Equations (\ref{4.14}) and (\ref{4.15}) are equivalent to Eqs. (86) and (87) of that paper, the first and second of Eqs. (\ref{4.18}) are equivalent to the second and first of Eqs. (88), respectively, and Eqs. (\ref{4.32}) and (\ref{4.33}) are equivalent to the first and second of Eqs. (89). These equivalences are manifestations of the isomorphism between Sp(2) and SU(1,1), which was discussed in \cite{mck25c}.

How do the SU(1,1) results compare to the SO(1,2) results? Equations (\ref{4.16}) and (\ref{4.18}) do not resemble Eqs. (\ref{3.22}) and (\ref{3.24}), but Eqs. (\ref{4.31}) -- (\ref{4.33}) do ($D_i \leftrightarrow \ga_i$ and $T_i \leftrightarrow u_i$). Furthermore, Eq. (\ref{4.34}) resembles Eq. (\ref{3.25}). The only difference between the SU(1,1) formulas and the SO(1,2) formulas is that the former involve double arguments, whereas the latter involve single arguments. This argument difference is caused by a normalization difference, which we now explain.

Most indefinite unitary matrices (the ones of practical interest) can be written as the exponentials of generating matrices: $M = \exp(G)$, where $G = \tsum_i G_ik_i$ is a linear combination of basis generators.  For SU(1,1), the basis generators are
\be G_1 = \left[\begin{array}{cc} 0 & 1 \\ 1 & 0\end{array}\right], \ \ 
G_2 = \left[\begin{array}{cc} 0 & i \\ -i & 0 \end{array}\right], \ \ 
G_3 = \left[\begin{array}{cc} i & 0 \\ 0 &- i \end{array}\right]. \label{4.41} \ee
These matrices satisfy the commutation relations
\be [G_1, G_2] = -2G_3, \ \ [G_2, G_3] = 2G_1, \ \ [G_3, G_1] = 2G_2, \label{4.42} \ee
where the commutator $[x, y] = xy - yx$. By using the identities $G_1^2 = G_2^2 = I$ and $G_3^2 = -I$, one can show that matrices (\ref{4.41}) generate the fundamental matrices (\ref{4.2}) and (\ref{4.3}). In this context, the generator coefficients $k_1 = \ze_1$, $k_2 = \ze_2$ and $k_3 = \ph_3$.

Likewise, every indefinite orthogonal matrix of practical interest can be written in the exponential form $L = \exp(H)$, where $H = \tsum_i H_il_i$. For SO(1,2), the generators
\be H_1 = \left[\begin{array}{ccc} 0 & 1 & 0 \\ 1 & 0 & 0 \\ 0 & 0 & 0 \end{array}\right], \ \ 
H_2 = \left[\begin{array}{ccc} 0 & 0 & 1 \\ 0 & 0 & 0 \\ 1 & 0 &0 \end{array}\right], \ \ 
H_3 = \left[\begin{array}{ccc} 0 & 0 & 0 \\ 0 & 0 & -1 \\ 0 & 1 & 0 \end{array}\right] \label{4.43} \ee
satisfy the commutation relations.
\be [H_1, H_2] = -H_3, \ \ [H_2, H_3] = H_1, \ \ [H_3, H_1] = H_2. \label{4.44} \ee
It is easy to verify that matrices (\ref{4.43}) generate the fundamental matrices (\ref{3.3}) and (\ref{3.4}). In this context, $l_1 = \ze_1$, $l_2 = \ze_2$ and $l_3 = \th_3$.
One relates the two sets of exponential matrices by relating their coefficients $k_i$ and $l_i$ (so the matrices have proportional coefficients, but different generators).

The generators of SU(1,1) satisfy commutation relations (\ref{4.42}), in which the structure coefficients are $\pm 2$, whereas the generators of SO(1,2) satisfy relations (\ref{4.44}), in which the coefficients are $\pm 1$ \cite{mck25c}. To make fair comparisons between the SU(1,1) and SO(1,2) results, one must replace the SU(1,1) parameters $\ze$, $\ph$ and $\th$ by $\ze/2$, $\ph/2$ and $\th/2$, respectively (where $\ph$ and $\th$ are the input and output phases). These replacements change the double arguments in Eqs. (\ref{4.31}) -- (\ref{4.34}) into single arguments. The modified versions of these equations are equivalent to Eqs. (\ref{3.22}) -- (\ref{3.25}). Thus, SU(1,1) is locally isomorphic to SO(1,2).

If one knew in advance that SU(1,1) is locally isomorphic to SO(1,2), and which unitary generator corresponds to which orthogonal generator, then one would only need to derive Eqs. (\ref{4.18}) and (\ref{4.22}), which is easy to do ($\th_w = \ph_\mu$). However, if one had to establish the iso-morphism first, for example by showing that the relations between the generator coefficients of the product and constituent matrices are equivalent \cite{mck25c}, then doing so, and using the SU(1,1) results to deduce the corresponding the SO(1,2) results, would require more work than deriving the SO(1,2) results directly.

In time and three space dimensions, Lorentz matrices are members of SO(1,3), which is locally isomorphic to the special linear group SL(2,C) \cite{jac99}. In principle, one can derive results for SL(2,C) and deduce the corresponding results for SO(1,3).

\newpage

\sec{5. Summary}

To practitioners of special relativity, it is well known that the composition of two nonparallel Lorentz boosts is not another boost: It is a boost followed by a rotation. Although it is straightforward to calculate the energy and direction of the combined boost, it is difficult to calculate the rotation angle, which is called the Wigner angle. In this article, we described three ways to determine the Wigner angle.

In Sec. 2, we used the vector (tensor) formalism to analyze the composition of two boosts.
Two equivalent formulas for the associated product tensor were derived. The first formula involves the momenta $\vu_1$ and $\vu_2$ that define the boosts [Eqs. (\ref{2.11}) -- (\ref{2.15})], whereas the second involves the momenta $\vu_{12}$ and $\vu_{21}$ that appear in the product tensor [Eqs. (\ref{2.21}) and (\ref{2.25})]. The derivation of the second formula was based on the {\it a priori} assumption that the composition of two boosts is a boost followed by a rotation.
We explained why the angle between $\vu_{12}$ and $\vu_{21}$ is the Wigner angle, and used the cross- and dot-products of these vectors to derive formulas for the sine and cosine of this angle [Eqs. (\ref{2.34}) and (\ref{2.35})].

In Sec. 3, we used the matrix formalism to analyze the same problem.
The special orthogonal group SO(1,2) consists of real $3 \times 3$ matrices $L$ that satisfy the indefinite orthogonality (Lorentz) condition $L^tSL = S$, where $S = \diag(1,-1,-1)$ is the structure (metric) matrix. This condition ensures that Lorentz transformations of the coordinate vector $T = [t, x, y]^t$ preserve the inner product (squared interval) $T^tST = t^2 - x^2 - y^2$.
The Lorentz condition imposes significant constraints on Lorentz matrices, which ensure that each matrix has the Schmidt-like decomposition $L(\ga, \th_1, \th_2) = R(\th_2)B(\ga)R^t(\th_1)$, where $B(\ga)$ represents a boost in the $x$ direction with energy $\ga$, and $R(\th_i)$ represents a rotation through the (input or output) angle $\th_i$ [Eqs. (\ref{3.11}) -- ({\ref{3.13})]. This decomposition can be rewritten in the equivalent form $L(\ga, \th_1, \th_2) = R(\th_{21})B(\ga,\th_1)$, where $B(\ga,\th_1) = R(\th_1)B(\ga)R^t(\th_1)$ represents a boost with direction angle $\th_1$ and $R(\th_{21})$ represents a rotation through the difference (Wigner) angle $\th_{21} = \th_2 - \th_1$. (The second decomposition justifies the assumption mentioned above.)
If one is presented with a Lorentz matrix, decomposition (\ref{3.12}) allows one to read off the energy [Eq. (\ref{3.14})], and provides equations for the input and output angles [Eqs. (\ref{3.15})]. It also provides equivalent equations for the cosine, sine and tangent of the Wigner angle [Eqs. (\ref{3.16}) -- (\ref{3.18})].
Rotating the axes used to analyze a Lorentz transformation corresponds to making a similarity transformation of the associated Lorentz matrix [Eq. (\ref{3.19})]. Such a transformation changes the input and output angles in simple ways, but does not change the Wigner angle or the energy. In our analysis of two boosts, we used rotated axes to simplify the calculation of the product matrix [Eq. (\ref{3.21})], then converted the results from the rotated frame back to the laboratory frame. The combined energy was specified in Eq. (\ref{3.22}), the input and output angles were specified in Eqs. (\ref{3.24}) and (\ref{3.26}), and the tangent of the Wigner angle was specified in Eq. (\ref{3.25}).

In Sec. 4, we demonstrated the local isomorphism (structural similarity) between SU(1,1) and SO(1,2). 
The special unitary group SU(1,1) consists of complex $2 \times 2$ matrices $M$ that satisfy the indefinite unitarity condition $M^\d SM = S$, where $S = \diag(1,-1)$ is the metric matrix. This condition ensures that indefinite unitary transformations of the spinor $A = [u, v]^t$ preserve the inner product $A^\d SA = |u|^2 - |v|^2$.
The indefinite unitarity condition imposes significant constraints on indefinite unitary matrices, which ensure that each matrix has the Schmidt-like decomposition $M(\mu, \ph_1, \ph_2) = P(\ph_2)B(\ga)P^t(\ph_1)$, where $B(\mu)$ represents a real boost with strength $\mu$, and $P(\ph_i)$ represents a differential phase shift through the (input or output) phase angle $\ph_i$ [Eqs. (\ref{4.7}) and ({\ref{4.9})].
Changing the reference phase of an indefinite unitary transformation corresponds to making a similarity transformation of the associated matrix [Eq. (\ref{4.4})]. Such a transformation changes the input and output phases in simple ways, but does not change the difference (Wigner) phase or the boost strength. The similarities between these results and the preceding ones are obvious.
The strength of a combined boost was specified in Eq. (\ref{4.31}), the input and output phases were specified in Eqs. (\ref{4.32}) and (\ref{4.33}), and the tangent of the Wigner half- and full-phases were specified in Eqs. (\ref{4.18}) and (\ref{4.34}). The equivalence of the indefinite unitary and orthogonal results is a manifestation of the local isomorphism between SU(1,1) and SO(1,2). For the problem considered (composition of boosts), the indefinite unitary and orthogonal analyses were of comparable difficulty.

Finally, for completeness, the composition of two arbitrary transformations is discussed in the appendix. The analysis of this general problem is no more difficult than that of the special problem described above.

In conclusion, we described three methods to determine the Wigner angle, all of which work well. Although the indirect (local isomorphism) method significantly simplifies the analysis of rotations in three dimensions \cite{mck25c,rui13}, it does not simplify the analysis of Lorentz transformations in time and two space dimensions, because the direct (matrix and vector) methods are straightforward to implement.

\newpage

\sec{Appendix: Two arbitrary transformations}

In this appendix, we generalize the analysis of Sec. 4, which applied to the composition of two boosts.
For two arbitrary SU(1,1) transformations,
\be Y = (Q_2B_2P_2^\d)(Q_1B_1P_1^\d)X. \label{a1} \ee
Equation (\ref{4.21}) can be written in the form $Y = M_4X$, where the final matrix $M_4 = Q_2M_3P_1^\d$ and the intermediate matrix $M_3 = B_2(P_2^\d Q_1)B_1$. In the last equation, $P_2^\d Q_1 = P^\d(\ph_{21})$, where $\ph_{21} = \ph_2 - \th_1$. Written explicitly,
\ba M_3 &= &\left[\begin{array}{cc} \mu_2 & \nu_2 \\ \nu_2 & \mu_2 \end{array}\right]
\left[\begin{array}{cc} e_{21}^* & 0 \\ 0 & e_{21} \end{array}\right]
\left[\begin{array}{cc} \mu_1 & \nu_1 \\ \nu_1 & \mu_1 \end{array}\right] \nonumber \\
&= &\left[\begin{array}{cc} \mu_2\mu_1e_{21}^* + \nu_2\nu_1e_{21} & \mu_2\nu_1e_{21}^* + \nu_2\mu_1e_{21} \\ \nu_2\mu_1e_{21}^* + \mu_2\nu_1e_{21} & \mu_2\mu_1e_{21} + \nu_2\nu_1e_{21}^* \end{array}\right], \label{a2} \ea
where $e_{21} = \exp(i\ph_{21})$.
The squared boost parameters are
\ba |\mu_3|^2 &= &\mu_2^2\mu_1^2 + \nu_2^2\nu_1^2 + 2\mu_2\nu_2\mu_1\nu_1(c_{21}^2 - s_{21}^2), \label{a3} \\
|\nu_3|^2 &= &\mu_2^2\nu_1^2 + \nu_2^2\mu_1^2 + 2\mu_2\nu_2\mu_1\nu_1(c_{21}^2 - s_{21}^2). \label{a4} \ea
It is easy to verify that the difference $|\mu_3|^2 - |\nu_3|^2 = 1$ and the sum
\be |\mu_3|^2 + |\nu_3|^2 = (\mu_2^2 + \nu_2^2)(\mu_1^2 + \nu_1^2)
+ (2\mu_2\nu_2)(2\mu_1\nu_1)(c_{21}^2 - s_{12}^2). \label{a5} \ee
The component phases are specified implicitly by the equations
\be \tan(\ph_{\mu3}) = -{(\mu_2\mu_1 - \nu_2\nu_1)s_{21} \over (\mu_2\mu_1 + \nu_2\nu_1)c_{21}}, \ \ 
\tan(\ph_{\nu3}) = {(\nu_2\mu_1 - \mu_2\nu_1)s_{21} \over (\nu_2\mu_1 + \mu_2\nu_1)c_{21}}. \label{a7} \ee
It follows from Eqs. (\ref{4.7}) and (\ref{a7}) that the input and output phases are specified by the equations
\ba \tan(2\ph_3) &= &{2\mu_2\nu_2s_{21}c_{21}
\over (\mu_2^2 + \nu_2^2)\mu_1\nu_1 + \mu_2\nu_2(\mu_1^2 + \nu_1^2)(c_{21}^2 - s_{21}^2)},  \label{a8} \\
\tan(2\th_3) &= &{-2\mu_1\nu_1s_{21}c_{21}
\over \mu_2\nu_2(\mu_1^2 + \nu_1^2) + (\mu_2^2 + \nu_2^2)\mu_1\nu_1(c_{21}^2 - s_{21}^2)}. \label{a9} \ea
The final parameters
\be \mu_4 = \mu_3e_2e_1^*, \ \ \ph_4 = \ph_3 + \ph_1, \ \ \th_4 = \th_3 + \th_2, \label{a10} \ee
from which it follows that the final difference (Wigner) phase 
\be \th_{w4} = \th_{w3} + \th_2 - \ph_1, \label{a11} \ee
where $\th_{w3} = \th_{\mu3}$ was specified by the first of Eqs. (\ref{a7}).

By using the double-argument notation of Sec. 4, one can rewrite Eqs. (\ref{a5}), (\ref{a8}) and (\ref{a9}) in the forms
\ba D_3 &= &D_2D_1 + T_2T_1d_{21}, \label{a12} \\
\tan(2\ph_3) &= &{T_2t_{21} \over D_2T_1 + T_2D_1d_{21}}, \label{a13} \\
\tan(2\th_3) &= &{-T_1t_{21} \over T_2D_1 + D_2T_1d_{21}}, \label{a14} \ea
respectively. By using the identities $\mu_2^2\mu_1^2 + \nu_2^2\nu_1^2 = (D_2D_1 + 1)/2$ and $\mu_2^2\mu_1^2 - \nu_2^2\nu_1^2 = (D_2$ $+ D_1)/2$, one can rewrite the first of Eqs. (\ref{a7}) in the double-argument form
\be \tan(2\ph_{\mu3}) = {-(D_2 + D_1)t_{21} \over T_2T_1 + (D_2D_1 + 1)d_{21}}. \label{a15} \ee

Notice that Eqs. (\ref{a12}) and (\ref{a13}) are identical to Eqs. (\ref{4.31}) and (\ref{4.32}), respectively, whereas Eq. (\ref{a14}) differs from Eq. (\ref{4.33}). These results reflect the fact that $M' = P_{21}M_3$: The matrices of interest only differ by an output phase shift, so the boost strengths and input angles are equal, whereas the output angles are different. The Wigner phases also differ, but in a relatively simple way ($\th'_w$ depends on $\ph_2 - \ph_1$ and $\th_w = \th'_w$, whereas $\th_3$ depends on $\ph_2 - \th_1$ and $\th_{w4} = \th_{w3} + \th_2 - \ph_1$). Overall, the general problem is no harder to solve than the special problem (upon which many papers have focused).

In passing, the composition of two symplectic transformations was discussed in \cite{mck25a}. Equations (\ref{a12}) -- (\ref{a14}) are identical to Eqs. (C9), (C11) and (C12) of that paper. These identities are manifestations of the isomorphism between Sp(2) and SU(1,1), which was discussed in \cite{mck25c}.

The combination of two Lorentz transformations was discussed in \cite{mck25b}. The input--output equation has the same form as Eq. (\ref{a1}), but the matrices involved are real $3 \times 3$ matrices (so $\d$ is replaced by $t$), $B_i$ represents a boost in the $x$ direction, and $P_i$ and $Q_i$ represent rotations about the $t$ axis (in the $xy$ plane).
The intermediate matrix
\ba L_3 &= &\left[\begin{array}{ccc} \ga_2 & u_2 & 0 \\ u_2 & \ga_2 & 0 \\ 0 & 0 & 1 \end{array}\right]
\left[\begin{array}{ccc} 1 & 0 & 0 \\ 0 & c & s \\ 0 & -s & c \end{array}\right]
\left[\begin{array}{ccc} \ga_1 & u_1 & 0 \\ u_1 & \ga_1 & 0 \\ 0 & 0 & 1 \end{array}\right] \nonumber \\
%&= &\left[\begin{array}{ccc} \ga_2 & u_2 & 0 \\ u_2 & \ga_2 & 0 \\ 0 & 0 & 1 \end{array}\right]
%\left[\begin{array}{ccc} \ga_1 & u_1 & 0 \\ cu_1 & c\ga_1 & s \\ -su_1 & -s\ga_1 & c \end{array}\right] \nonumber \\
&= &\left[\begin{array}{ccc} \ga_2\ga_1 + u_2u_1c & \ga_2u_1 + u_2\ga_1c & u_2s \\ u_2\ga_1 + \ga_2u_1c & u_2u_1 + \ga_2\ga_1c & \ga_2s \\ -u_1s & -\ga_1s & c\end{array}\right], \label{a21} \ea
where $c = \cos(\ph_2 - \th_1)$ and $s = \sin(\ph_2 - \th_1)$. This matrix is the product of Lorentz matrices, so it is also a Lorentz matrix, with the decomposition $L_3 = Q_3(\th_3)B_3(\ga_3)P_3^t(\ph_3)$.

It follows from Eqs. (\ref{3.14}) and (\ref{a21}) that the intermediate energy
\be \ga_3 = \ga_2\ga_1 + u_2u_1c, \label{a22} \ee
which is a symmetric function of $\ga_1$ and $\ga_2$, and depends on the difference angle $\ph_2 - \th_1$.
%(If the order of the transformations were reversed, it would depend on the angle $\ph_1 - \th_2$.) 
According to Eq. (\ref{3.12}), the first row of the product matrix is $[\ga_3, u_3c_\ph, u_3s_\ph]$, where $c_\ph = \cos(\ph_3)$, and the first column is $[\ga_3, u_3c_\th, u_3s_\th]^t$, where $c_\th = \cos(\th_3)$. The definitions of $s_\ph$ and $s_\th$ are similar.
It follows from Eqs. (\ref{3.15}) and (\ref{a21}) that the intermediate input and output angles are specified implicitly by the equations
\be \tan(\ph_3) = {u_2s \over \ga_2u_1 + u_2\ga_1c}, \ \ 
\tan(\th_3) = {-u_1s \over u_2\ga_1 + \ga_2u_1c}. \label{a24} \ee
It follows from Eqs. (\ref{3.18}) and (\ref{a21}) that the intermediate difference angle is specified by the equation
\be \tan(\th_3 - \ph_3) = {-(\ga_2 +\ga_1)s \over u_2u_1 + (\ga_2\ga_1 + 1)c}. \label{a25} \ee
If one were to derive Eq. (\ref{a25}) directly, by using Eqs. (\ref{a24}), one would obtain a fraction whose numerator and denominator are both proportional to $\ga_3 - 1$. By canceling this common factor, one would obtain Eq. (\ref{a25}).

The final matrix $L_4 = Q_2(\th_2)L_3(\ga_3,\ph_3,\th_3)P_1^t(\ph_1)$, from which it follows that
\be \ga_4 = \ga_3, \ \ \ph_4 = \ph_1 + \ph_3, \ \ \th_4 = \th_2 + \th_3. \label{a26} \ee
It also follows that
\be \th_4 - \ph_4 = \th_3 - \ph_3 + \th_2 - \ph_1. \label{a27} \ee
Equations (\ref{a22}) -- (\ref{a27}) are valid for arbitrary transformations (combinations of boosts and rotations).
They are the product rules for Lorentz matrices, written in terms of the Schmidt-like parameters $\ga$, $\ph$ and $\th$.
In particular, Eqs. (\ref{a25}) and (\ref{a27}) specify the difference (Wigner) angle. Given the complexity of matrices of the form (\ref{3.12}), it is remarkable that the product rules and the difference formula are so simple.

As we explained in the text, the generators of SU(1,1) and SO(1,2) have different normalizations. To make a fair comparison between results for SU(1,1) and SO(1,2), one must replace the SU(1,1) coefficients $\ze$, $\ph$ and $\th$ by $\ze/2$, $\ph/2$ and $\th/2$, respectively. These replacements change the double arguments in Eqs. (\ref{a12}) -- (\ref{a14}) into single arguments, like the ones in Eqs. (\ref{a22}) and (\ref{a24}). The SU(1,1) and SO(1,2) equations are equivalent, because the groups are locally isomorphic.
Although one could use this relation to deduce the SO(1,2) formulas from the SU(1,1) formulas, the SO(1,2) calculation is straightforward by itself, so no deduction is necessary.

\end{document}